\documentclass[referee]{raa}            % referee version: for submission

\usepackage{graphicx,times}             %for PS/EPS graphics inclusion, new
\usepackage{booktabs}
\usepackage{multirow}
\usepackage{amsmath}
\usepackage{natbib}
\usepackage{amssymb,amsmath}
\usepackage[dvipsnames]{xcolor}
\bibpunct{(}{)}{;}{a}{}{,}

\newcommand{\adr}[1]{\textcolor{red}{ #1}} % additions in red
\usepackage[pagebackref=true]{hyperref}

\begin{document}

  \title{Comparing galaxy merger orbits in hydrodynamical simulation and in dark-matter-only simulation}
%   \subtitle{I. Place Your Subtitle Here}

   \volnopage{Vol.0 (20xx) No.0, 000--000}      %%preserved for Editor. DOn't remove!
   \setcounter{page}{1}          %%starting page, preserved for Editor. DOn't remove!

   \author{Yahan Pu %(周爱英) %% Put your Chinese name in "( )" if you like. Note to open line 11 "\usepackage[UTF8]{ctex}"
      \inst{1,2}
   \and Lan Wang
      \inst{1,3}
   \and Guangquan Zeng
      \inst{4}
      \and Lizhi Xie
      \inst{5}
   }

   \institute{National Astronomical Observatories, Chinese Academy of Sciences,
             Beijing 100012, China; {\it yahanpu.astro@gmail.com; wanglan@bao.ac.cn}\\
        \and
             School of Mathematics and Physics, University of Science and Technology Beijing 100083, China; \\
        \and
            School of Astronomy and Space Science, University of Chinese Academy of Sciences, Beijing 100049, China;\\
        \and
            Department of Physics, The Chinese University of Hong Kong, Sha Tin, N.T., Hong Kong, China;\\
        \and
            Astrophysics Center, Tianjin Normal University, Xiqing, Tianjin 300387, China\\
\vs\no
   {\small Received 20xx month day; accepted 20xx month day}}

\abstract{
To investigate how the presence of baryons in simulations affects galaxy merger orbits, we compare in detail the merger timescales and orbits of the matched merger pairs in TNG100 hydrodynamical simulations and their corresponding dark-matter-only simulations, for different resolution levels. Compared with the mergers in the TNG100-1-Dark simulation without baryons, the matched mergers in the TNG100-1 simulation have similar infall time, but have statistically earlier merger times and therefore shorter merger timescales. The merger orbits for the matched pairs in the TNG100-1 and the TNG100-1-Dark simulations are similar right after infall, and both evolve to more head-on orbits at final stages, with smaller changes in the hydrodynamical simulation. In the final 2 Gyr before merger, the collision angles that represent merger orbits quantitatively are smaller in TNG100-1 than those in TNG100-1-Dark, by around 6$^\circ$ to 10$^\circ$, depending on the mass ratios and galaxy masses investigated. Our results demonstrate that the presence of baryons slightly accelerates the merger processes, and results in more spiral-in orbits for both major and minor mergers in galaxies with various stellar masses. These effects are less obvious in simulations with lower resolutions. 
}

   \authorrunning{Y. Pu,  et al. }            %author_head in even pages
   \titlerunning{Comparing galaxy merger orbits}  % title_head in odd pages

   \maketitle
%% The author head (on even pages) and the title head (on odd pages) will be
%% automatically extracted from \author{} and \title{}. Whenever the title is too long,
%% you will be asked to supply a shorter one by inserting either \authorrunning{} or
%% \titlerunning{} before \maketitle. Anyway, you can specify your own heads.
%%
%%
%% Note: In the following text body of your manuscript, please note several differences from
%%       other major journals:
%% (1) \subsection{Please Capitalize the First Letter of Each Notional Word in Subsection Title}
%% (2) Please Capitalize the First Letter of Each Notional Word in all tables' captions

%
%________________________________________________ sections below
%
\section{Introduction\label{intro}}           %% first-level sections will be auto-capitalized
\label{sect:intro}
%merger is important
Galaxy mergers play a crucial role in galaxy evolution, closely related to starbursts, black hole growth, and transformation of galaxy morphologies \citep{toomre1972galactic,1996ApJ...464..641M,hopkins2006unified}. Observationally, galaxy mergers are identified via signatures such as close galaxy pairs, tidal features, double nuclei, shells, and strong morphological asymmetries \citep[e.g.,][]{conselice2003direct,ryan2008galaxy}. However, while a galaxy merger is a process that can span several gigayears, for each merger event, what we observe is essentially just one snapshot of the merger in progress \citep{lotz2008galaxy,kitzbichler2008calibration}.
On the other hand, models and especially numerical simulations are usually used to study galaxy mergers, to obtain a continuous view of the whole merger process and to investigate the individual and statistical impact of mergers on the evolution of galaxies \citep[e.g.,][]{lotz2011major,lopez2013massiv,peschken2020disc,fuentealba2025deep}.

%Merger affects galaxy morphology.
Galaxy morphologies are expected to be significantly affected by mergers, especially major mergers. The general picture is that major mergers produce elliptical galaxies \citep{toomre1977theories,white1978core}, and minor mergers, although less disruptive, can still thicken galactic disks and contribute to 
the growth of the bulge component
%an increased population of elliptical galaxies 
\citep{1996ApJ...460..121W,bournaud2007multiple}. Therefore, galaxies at low redshift are more likely to be elliptical, as they are expected to have undergone more mergers according to the standard hierarchical model of galaxy formation \citep{gonzalez2005elliptical,kormendy2009structure}. This is consistent with the relative fraction of elliptical and disk galaxies observed in the local and high-redshift Universe \citep{buitrago2008size,retzlaff2010great,van20143d,conselice2014evolution}.
However, major mergers have also been found to be able to produce disk galaxies \citep{springel2005formation,sparre2017unorthodox,peschken2020disc}. Therefore, the relation between mergers and galaxy morphology transformation is still not completely clear. 

Numerical simulations have demonstrated that various merger properties significantly influence the morphology of the remnant galaxy, such as the merger mass ratio, the gas fraction of merger, the orbital configuration of merger, and so on.
For instance, the extensively-studied mass ratio has been shown to be important.
Using idealized high-resolution simulations, \citet{bournaud2005galaxy} and \citet{rodriguez2017role} demonstrated that major mergers (with mass ratios close to $1:1$) typically destroy stellar disks and form spheroids, whereas minor mergers are more likely to preserve the disk structures.
Moreover, gas-rich mergers appear more likely to preserve or even regenerate a disk structure in the merger remnant compared to gas-poor mergers \citep[e.g.,][]{2009ApJ...691.1168H, peschken2020disc, 2020MNRAS.494.5568J}.
Furthermore, the orbital configuration of the merging system also exerts a non-negligible influence on the resulting galaxy morphology \citep[e.g.,][]{2018MNRAS.480.2266M, 2022MNRAS.509.5062L, 2024RAA....24g5019H}.

% Numerical simulations have demonstrated that various merger properties affect
% %can significantly affect both the outcome of mergers and the properties of remnant galaxies
% the morphology of remnant galaxies. For the most studied mass ratio, for example, \citet{bournaud2005galaxy} and \citet{rodriguez2017role} 
% used idealized high-resolution simulations to show that major mergers with mass ratios close to 1:1 tend to destroy disks and form spheroids, whereas minor mergers are more likely to preserve disk structures.
% %\citet{rodriguez2017role}, using the cosmological Illustris simulation, found similar trends and emphasized the role of the large-scale environment in shaping galaxy morphology. \adb{(large-scale environment? not properties of mergers?)}
% Other properties investigated include cold gas fraction and orbital configuration (e.g. prograde versus retrograde). For example, \citet{lotz2010effect} found that higher gas fractions in disk galaxy mergers lead to stronger and more persistent morphological disturbances.
% %\adr{the visible shape irregularities in a galaxy, with orbital parameters(e.g. pericentric distance, orbital eccentricity, and disc–orbit orientation) playing a secondary role. Asymmetry mainly flags major mergers, for minor mergers the asymmetry window is short.}
% \adb{(GQ: add one or two more examples regarding cold gas fraction, orbital parameters - with different parameters describing orbits?, and any other merger property?)}

In a recent study by \citet{zeng2021formation}, for massive galaxies that experience major mergers, orbital type is found to be the key factor that determines the post-merger galaxy morphology, outweighing previously investigated properties such as cold gas fraction and orbital configuration. Specifically, the orbital type of mergers is represented by the collision angle, defined as the average acute angle between the relative position vector and the relative velocity vector of the satellite galaxy with respect to the central galaxy, measured from 1 Gyr before merger till merger time.
%over the merging process (from 1 Gyr before the merger until the merger occurs). 
With the IllustrisTNG simulation \citep{pillepich2018simulating}, \citet{zeng2021formation} showed that major mergers with a spiral-in orbit (i.e., large collision angle) mostly lead to disk-dominant remnants, while major mergers of head-on collision (i.e., small collision angle) mostly form ellipticals. 

%They showed that, in the IllustrisTNG simulation, the morphology of massive galaxy remnants depends sensitively on the orbit type of major mergers : spiral-infall orbits with large collision angle favor disc survival, while head-on collisions with small collision angle typically result in elliptical remnants. Notably, this orbital dependenceappears more significant than that of cold gas content or other orbital parameters. 

While based on the IllustrisTNG simulation, it is not clear whether the results of \citet{zeng2021formation} hold for other hydrodynamical simulations with different descriptions of baryon processes
\citep{Schaye_2007,schaye2015eagle,dubois2016horizon,dave2019simba,zana2022enhanced}.
%For example, stellar feedback, star formation, and AGN feedback models are employed in various ways in simulations of Eagle  \citep{Schaye_2007,schaye2015eagle}, Horizon-AGN \citep{dubois2016horizon}, and SIMBA \citep{dave2019simba}. 
These various treatments in baryonic processes would result in different galaxy merger rates \citep{hopkins2010mergers}, merger timescales \citep{xu2025describing}, positions and velocities of satellite galaxies \citep{contreras2025effectbaryonspositionsvelocities}, and may affect the orbital type, as well as the dependence of the galaxy morphology on the orbital type.
Apart from hydrodynamical simulations, other models of galaxy formation and evolution such as semi-analytic models and halo-based models, are built on dark-matter-only simulations, where galaxy evolution is linked to the merger histories of dark matter haloes/subhaloes. %without baryons, rather than the merger histories of galaxies themselves. 
In these models, the galaxy morphology is normally assumed to change from disky to elliptical when a major merger happens \citep{wang2019comparing, guo2011, Lacey2016, xie2017, stevens2024}%citep{Xie} \adb{(Xie Lizhi can add some classical papers here)}
, without considering the effect from orbital type. Before applying the dependence of merger remnant morphology on orbital type in a model that studies galaxy morphology evolution in more detail (Xie et al. in preparation), it is necessary to check how orbital type, and quantitatively how much the collision angle is affected by the presence of baryons in simulation. 
%Without baryons, the merger orbital type could be different for these pure dark matter halo mergers. Therefore, it is essential to check how the orbit type, and quantitatively how much the collision angle is affected by baryonic effect.

In this study, we compare in detail the merger orbits in matched merger samples of hydrodynamical TNG simulations and the corresponding dark-matter-only TNG-Dark simulations, to investigate how the existence of baryons affects the orbital types of galaxy mergers. 
In addition, while in simulations numerical resolution is a critical factor influencing the identification of structures and merger dynamics \citep{knebe2011haloes, behroozi2012rockstar, onions2012subhaloes}, we also compare mergers in TNG simulations of different resolutions to study the effects of resolution on merger orbital type.
% High-resolution simulations are better able to resolve satellite galaxies, especially those near the host halo center, thereby improving merger completeness and affecting the classification of orbital configurations \citep{knebe2011haloes, behroozi2012rockstar, onions2012subhaloes}. 
By comparing mergers in simulations with and without baryons, and across different resolution levels, we aim to 
%identify the physical drivers behind variations in 
understand better merger orbital type represented by the collision angle, the critical yet underexplored parameter in shaping galaxy morphology during mergers.

This paper is organized as follows.  Section~\ref{sect:Simulations} introduces the TNG Simulations used, and describes how merger samples and matched merger samples between hydrodynamical (hydro) and dark-matter-only (DMO) simulations are selected. Section~\ref{sect:Orbit type comparison} compares in detail the merger properties, especially the orbital types in TNG100-1 and TNG100-1-Dark simulations, for galaxies with various masses. In Section~\ref{section: Comparison in different resolutions}, we check the effect of resolution on orbital type using TNG simulations of different resolution levels. Discussion and conclusions are presented in Section~\ref{section: Discussion and conclusions}.

\section{Simulations and sample selection}
\label{sect:Simulations}
\subsection{IllustrisTNG simulations}
The IllustrisTNG project \citep{springel2018first, nelson2018first, naiman2018first, marinacci2018first, pillepich2018first} is a suite of cosmological galaxy formation simulations, performed with the moving-mesh code \texttt{AREPO} \citep{springel2010pur}.
Notably, IllustrisTNG has demonstrated its ability to reproduce the observed statistical properties of galaxy morphology \citep[e.g.,][]{tacchella2019morphology}.
% As illustrated by \citep{tacchella2019morphology}, IllustrisTNG is able to reproduce the statistical properties of the galaxy morphology in good agreement with observations.

In this work, we utilize the publicly available data of the hydrodynamical simulation TNG100-1 and its dark-matter-only (DMO) counterpart TNG100-1-Dark to investigate the impact of baryons on galaxy merger orbit in detail.
To study the impact of resolution effect on the results, we also analyze lower-resolution simulations TNG100-2, TNG100-3, and their DMO counterparts TNG100-2-Dark, TNG100-3-Dark in Section~\ref{section: Comparison in different resolutions}.
It is noted that, each of these simulates a box with a side length of 75~\(h^{-1}\)~Mpc.
Moreover, the TNG100-1 contains \(1820^3\) dark matter particles and \(1820^3\) gas cells initially, with mass resolutions of \(m_{\mathrm{DM}} = 7.5 \times 10^{6} \, h^{-1} M_\odot\) and \(m_{\mathrm{gas}} = 1.4 \times 10^{6} \, h^{-1} M_\odot\),
and the TNG100-1-Dark contains only \(1820^3\) dark matter particles, with a mass resolution of \(m_{\mathrm{DM}} = 8.9 \times 10^{6} \, h^{-1} M_\odot\). 
For lower resolution runs, TNG100-2 has mass resolutions of \(m_{\mathrm{DM}} = 6.0 \times 10^{7} \, h^{-1} M_\odot\), \(m_{\mathrm{gas}} = 1.1 \times 10^{7} \, h^{-1} M_\odot\), and TNG100-2-Dark has \(m_{\mathrm{DM}} = 6.9 \times 10^{7} \, h^{-1} M_\odot\).
Similarly, TNG100-3 has \(m_{\mathrm{DM}} = 4.8 \times 10^{8} \, h^{-1} M_\odot\), \(m_{\mathrm{gas}} = 8.9 \times 10^{7} \, h^{-1} M_\odot\), and TNG100-3-Dark has \(m_{\mathrm{DM}} = 5.5 \times 10^{8} \, h^{-1} M_\odot\). 
%The comparison between these two runs, which share identical cosmology, box sizes, and initial conditions but differ in the inclusion of baryonic processes such as gas cooling, star formation, and feedback, allows us to isolate the effects of baryons on mergers.

%To further examine the accuracy of our results, we additionally make use of the lower-resolution simulations TNG100-2 and TNG100-3, along with their respective DMO counterparts TNG100-2-Dark and TNG100-3-Dark. These simulations enable a systematic analysis of how numerical resolution influences the derived merger statistics and the degree to which our conclusions hold across varying mass and spatial resolution. The TNG100-2, and TNG100-3 simulations also share a box size of 75~\(h^{-1}\)~Mpc but differ from TNG100-1 in resolution. 
%The TNG100-2 simulation has \(910^3\) dark matter particles and \(910^3\) gas cells, with dark matter and gas mass resolutions of \(6.0 \times 10^{7} \, h^{-1} M_\odot\) and \(1.1 \times 10^{7} \, h^{-1} M_\odot\), respectively. The TNG100-3 simulation, the lowest resolution in this suite, includes \(455^3\) dark matter particles and \(455^3\) gas cells. Its mass resolutions are \(4.8 \times 10^{8} \, h^{-1} M_\odot\) for dark matter and \(8.9 \times 10^{7} \, h^{-1} M_\odot\) for gas. The TNG100-2-Dark and TNG100-3-Dark simulations share the same 75~\(h^{-1}\)~Mpc box, containing \(910^3\) and \(455^3\) dark matter particles, with mass resolutions of \(6.9 \times 10^{7} \, h^{-1} M_\odot\) and \(5.5 \times 10^{8} \, h^{-1} M_\odot\), respectively.

In these simulations the dark matter haloes and subhaloes are identified using the \texttt{FoF} and \texttt{Subfind} algorithms \citep[][]{2001MNRAS.328..726S, 2009MNRAS.399..497D} respectively.
For hydrodynamical simulations, subhaloes with stellar components are considered to be galaxies.
The merger trees of subhaloes/galaxies are constructed using the \texttt{Sublink} algorithm \citep{rodriguez2015merger}. 
Based on the merger trees, a merger event is identified when two distinct subhaloes/galaxies share the same descendant.
%By tracing a subhalo backward through snapshots, its main progenitor branch is obtained, with the main progenitor at each snapshot defined as the most massive progenitor. 
%As galaxies are hosted by subhaloes, this approach allows us to recover their formation and assembly histories from the corresponding merger trees. The same method is applied consistently to both the hydrodynamical and the corresponding dark matter–only simulations.

\subsection{Matched merger samples of hydrodynamical and dark-matter-only simulations \label{merger samples}}
%check mass and mass ratio compare with previous work refs
To make a detailed comparison of merger orbits in simulations with and without baryons, we select one-to-one matched merger pairs from the TNG hydro simulation and its DMO counterpart at the same resolution.

In the hydrodynamical simulation, we firstly select galaxy mergers of different mass ratios based on the merger histories of galaxies with various stellar masses at $z = 0$. Following \citet{zeng2021formation}, we focus on the latest (and at $z<1$) major merger and minor merger of each galaxy investigated. 
To minimize the impact of external perturbations, we additionally exclude mergers that experience another major merger within 1~Gyr before or after the event.
Because a substantial fraction of the stellar mass of the satellite galaxy can be stripped during a merger \citep{wang2019comparing, lokas2020interesting}, we follow previous studies \citep{rodriguez2017role,eisert2023ergo} and calculate the merger mass ratio at the snapshot where the satellite attains its maximum stellar mass, while the two galaxies are still relatively isolated.
Note that this mass ratio calculation differs slightly from that in \citet{zeng2021formation}, where the mass ratio is calculated using the stellar masses of the two merging galaxies of their own maximums before merger.
Besides, the stellar mass of galaxies used in \citet{zeng2021formation} is the total mass of stellar particles contained within twice the stellar half-mass radius of the galaxy, while in this work we define stellar mass as the total mass of all stellar particles contained in the subhalo.
We have checked that, with the slight differences applied above, the main results of \citet{zeng2021formation} remain unchanged, with only minor variations in detail (see Fig. \ref{fig:compare with previous work} in Appendix).

For each merger selected in the hydrodynamical simulation, we then identify its one-to-one matched counterpart in the DMO simulation using the bidirectional matching catalog provided by the IllustrisTNG team \citep{rodriguez2015merger, nelson2015illustris}, which matches subhalos between different simulation runs. To build up the bidirectional matching catalog of subhalos, for
each subhalo in the hydro simulations at each snapshot, all subhalos in the DMO counterpart that share common dark-matter (DM) particle IDs are considered as potential matches, and the best match is chosen as the one with the largest number of shared DM particles. The same procedure is then performed in the reverse direction, to find the best match in the hydro counterpart for each subhalo in the DMO simulations. A matching is accepted only if the two matched subhaloes both find the other one as its best match in the corresponding simulation.
%If this bijective condition is not satisfied, it does not count as a match.
In this work, based on the bidirectional matching subhalos, we first select the matched galaxies/subhalos at z=0. Then, in their merger histories, the one-to-one merger pairs are included in our analysis only if the pre-merger central and satellite galaxies/subhaloes and the post-merger galaxy/subhalo in mergers selected from the hydro simulation all have matched counterparts in the DMO simulation.
For the matched merger pairs identified in the hydrodynamical and DMO simulations, the satellite subhalo in a few cases has a very low mass, corresponding to a small number of particles,
which could introduce significant error into our analysis.
We therefore only consider mergers where the DMO satellite subhalo comprises a minimum of 1,000 dark matter particles (i.e., \(8.9 \times 10^{9} \, h^{-1} M_\odot\) for TNG100-1) at its maximum mass. In this work, following \citet{zeng2021formation}, galaxy mergers identified in the hydrodynamical simulations are classified as major for stellar mass ratios exceeding 1 : 4, and as minor for ratios between 1 : 10 and 1 : 4. Focusing on galaxy stellar mass rather than halo mass, and  defining merger types based on stellar mass ratio rather than total mass ratio has the advantage of being able to be compared with observations more directly \citep{robotham2014galaxy,davies2015galaxy,duncan2019observational,li2023subtle}. 
Besides, stellar mass and the corresponding ratio identify the tightly bound material in galaxies, which actually induces structural perturbations and bulge formation that determine galaxy morphology \citep{stewart2009galaxy,hopkins2010mergers,rodriguez2015merger}.

Following \citet{wang2019comparing}, we divide our sample into three sub-samples based on galaxy stellar mass at $z = 0$ in the hydro simulation: (1) the Milky Way-mass galaxies (MW) with stellar masses of \( 4-8 \times 10^{10} \, \text{M}_{\odot} \); (2) less massive galaxies (Less) with \( 1-4 \times 10^{10} \, \text{M}_{\odot} \); and (3) more massive galaxies (Massive) with stellar masses greater than \( 8 \times 10^{10} \, \text{M}_{\odot} \).
Table~\ref{tab:examples} presents the number of selected mergers in TNG100-1 and the corresponding matched pairs between TNG100-1 and TNG100-1-Dark, for both major and minor mergers in these stellar mass bins.
% The numbers of selected mergers in TNG100-1, and the number of matched merger pairs of TNG100-1 and TNG100-1-Dark are shown in Table~\ref{tab:examples}, for major and minor mergers that happen in galaxies with different stellar masses, respectively.

%After applying the selection steps described above, these subsamples together contain a total of 3,582 mergers, as shown in Table~\ref{tab:examples}.

\renewcommand{\arraystretch}{1.2}
\begin{table}[h]
\centering
\caption{The total numbers of major and minor merger events selected in TNG100-1, for galaxies of different stellar masses at present day. The numbers in brackets are the numbers of mergers after one-to-one matching with the TNG100-1 DMO simulation.}
\label{tab:examples}
\begin{tabular}{lccc}
\toprule
 & Massive  & MW & Less  \\
\midrule
Major & 497 (259) & 412 (218) & 976 (491) \\
Minor & 430 (219) & 365 (202) & 902 (518) \\
\bottomrule
\end{tabular}
\end{table}

%\subsection{Matching mergers in DMO simulation}
%\label{Matching}
%To study in detail the difference between merger orbit in hydro and DMO simulations, and to investigate the effect of baryons on merger orbit in TNG, we apply one-to-one match between merger events in the TNG100-1 and TNG100-1-Dark simulation.

%Based on the merger examples selected in TNG100-1, we match corresponding merger events in the DMO simulations based on the following steps. We first perform a $z=0$ galaxy matching between the hydrodynamical and DMO runs using the LHaloTree bidirectional matching catalog \citep{rodriguez2015merger,nelson2015illustris}, retaining only those galaxies with a valid counterpart in the other simulation. For each matched galaxy in the hydro run, we select merger events from our pre-selected hydro merger sample. We then require that the secondary galaxy at infall also has a matched counterpart in the DMO run \citep{fakhouri2010dark}. This matched DMO galaxy must be a progenitor of the secondary galaxy involved in a merger of the DMO counterpart. Also, to ensure that mergers in the DMO simulation are well resolved, we require that satellite galaxies, at their maximum mass, contain at least 1,000 dark matter particles (\(8.9 \times 10^{9} \, h^{-1} M_\odot\)). Mergers satisfying these conditions are retained as the final matched sample. The matching DMO mergers were assigned the same classification as their corresponding hydro mergers. Using this approach, we identify 1,907 matched pairs out of 3,582 in in TNG100-1-Dark, as shown in Table~\ref{tab:examples}.

\section{Comparison of mergers in hydrodynamical and dark-matter-only simulations }
\label{sect:Orbit type comparison}

In the hydrodynamical simulations, baryons contribute a small fraction to the total mass of galaxies/subhalos. For example, for the matched massive galaxies in TNG100-1, the 10\%, 50\%, 90\% values of the distributions of baryon mass fraction and stellar mass fractions are 0.055, 0.084, 0.125 and 0.013, 0.027, 0.059, respectively. Nevertheless, baryons dominate in the central part of galaxies \citep{stewart2009galaxy,rodriguez2015merger,nevin2023declining}.
During mergers, the extended outer part of the satellite subhalo dominated by dark matter component is firstly stripped and/or mixed with the central halo, before the tightly bound, baryon dominated parts start to interact and cause galaxy morphological change \citep{2009ASPC..419..243S,hopkins2010mergers}. 

Compared with DMO simulation, the existence of baryons in hydro simulation would affect the properties of mergers, in terms of both time related and orbit type related. For example, ram pressure, which removes gas from satellites in groups and clusters \citep{rodriguez2022satellite,kulier2023ram,zhu2023and,chen2024environmental}, can cause a longer analytic dynamical-friction timescale with reduced satellite mass, but at the same time gas drag and orbit circularization can accelerate orbital decay for some massive satellites \citep{mccarthy2008ram}.
In addition, baryonic condensation can make satellites more tightly bound and resistant to tidal stripping, affecting merger times \citep{montero2024tracking,xu2025describing}.

In Section \ref{subsection orbits}, we compare the detailed merging orbit of the matched major mergers in massive galaxies, and extend the analysis to minor mergers, as well as to galaxies in other mass bins in Section \ref{Collision angle distribution for less massive galaxies and for minor mergers}.
%We also present the collision angle distributions in the two simulations. 

\subsection{Merger time comparison\label{Merger time comparison}}

Before looking into the detailed merger orbits of galaxy mergers, we first look at and compare the statistical differences in times related to mergers, for the matched merger pairs of TNG100-1 and TNG100-1-Dark.
For each merger, we record the infall time $t_{\mathrm{infall}}$, the merger time $t_{\mathrm{merger}}$, and then calculate the merger timescale $t_\mathrm{infall}-t_\mathrm{merger}$ afterwards.

Specifically, $t_{\mathrm{infall}}$ is defined as the lookback time of the first snapshot when the satellite galaxy/subhalo enters within $R_{200}$ (defined as the comoving radius within which the mean density is 200 times the critical density of the Universe) of the host halo of the central galaxy. $t_{\mathrm{merger}}$ is defined as the lookback time of the last snapshot in which the two galaxies can still be identified as separate galaxies before they finally merge into a single system.

\begin{figure}[htbp]
\centering
\begin{minipage}[t]{0.35\linewidth}
  \centering
  \includegraphics[height=0.18\textheight]{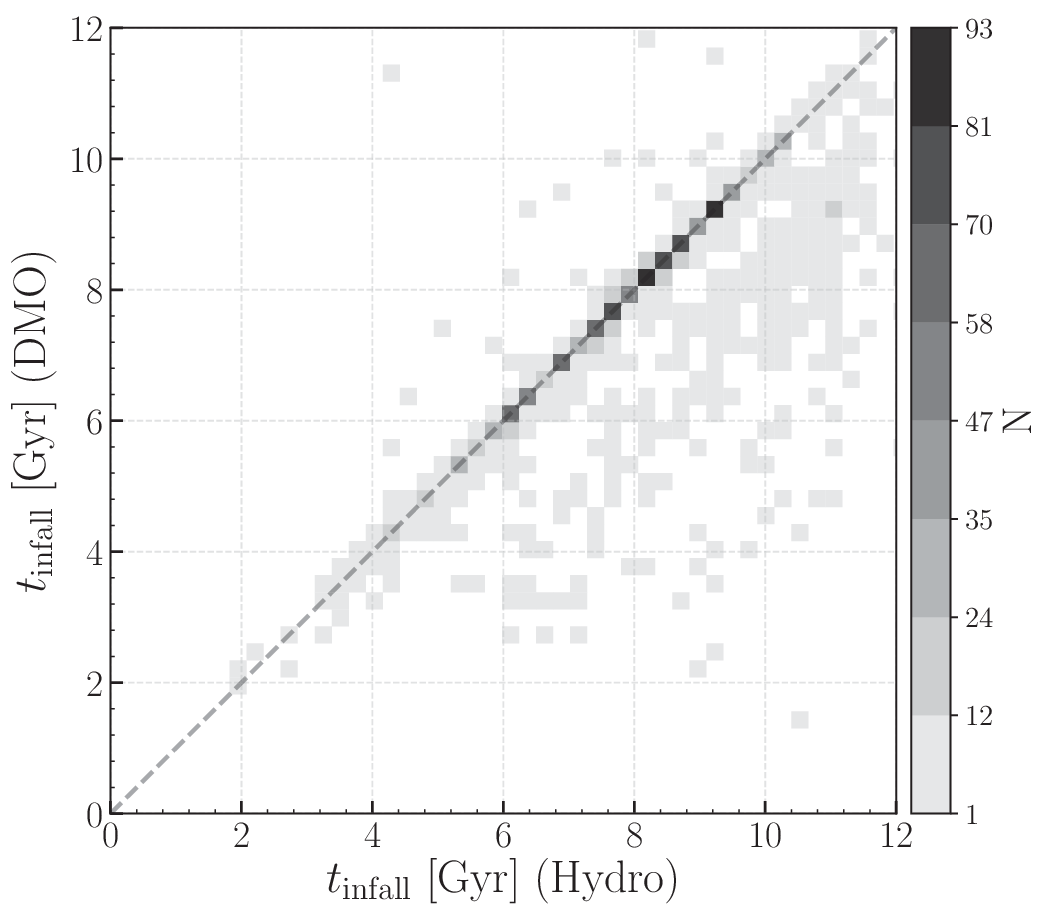}
\end{minipage}%
\begin{minipage}[t]{0.35\linewidth}
  \centering
  \includegraphics[height=0.18\textheight]{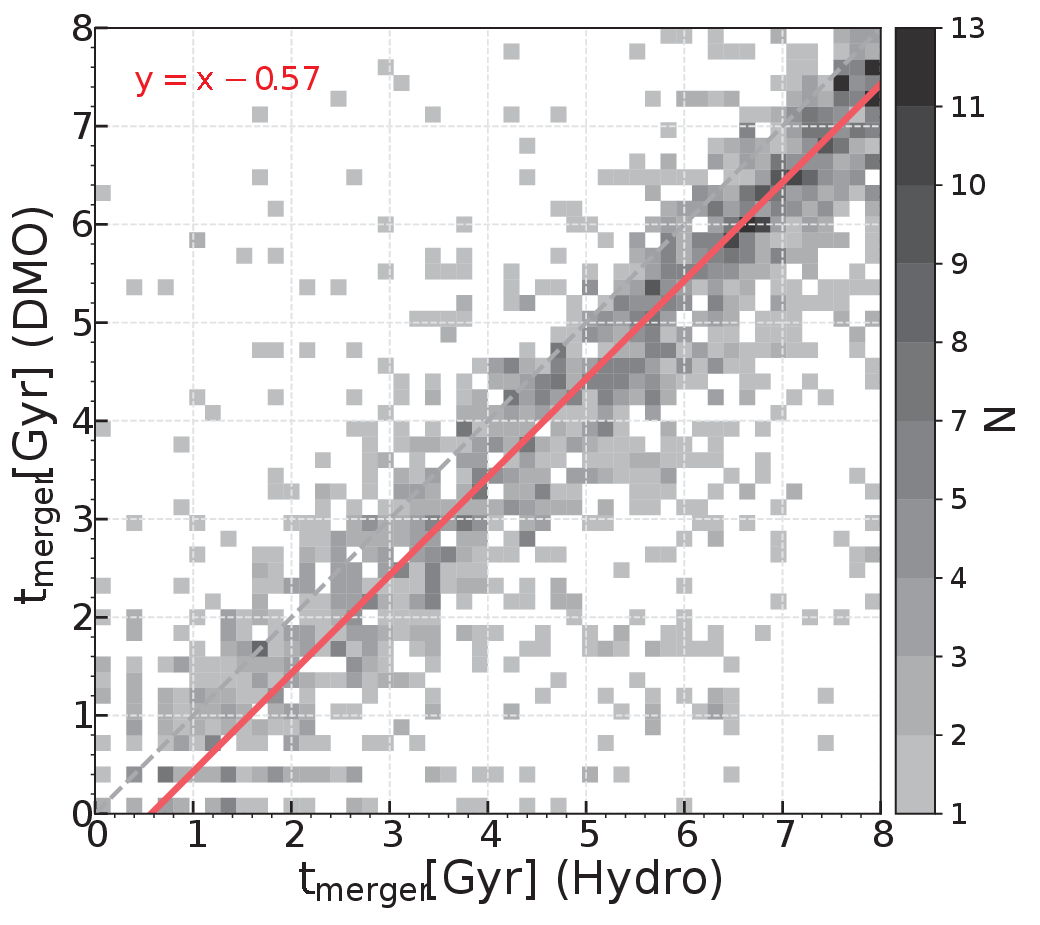}
\end{minipage}%
\begin{minipage}[t]{0.30\linewidth}
  \centering
  \includegraphics[height=0.18\textheight]{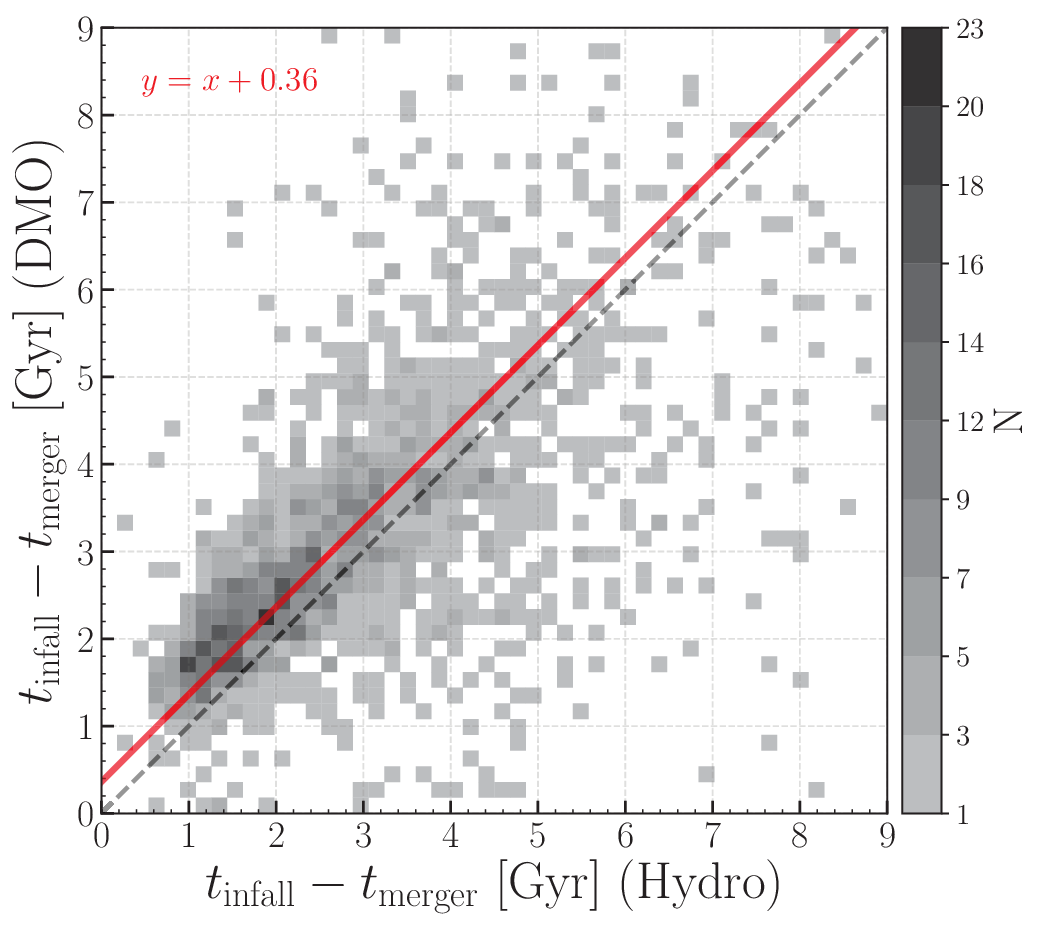}
\end{minipage}
\caption{Comparison of infall time ($t_{\mathrm{infall}}$, left panel), merger time ($t_{\mathrm{merger}}$, middle panel), and merger timescale ($t_{\mathrm{infall}} - t_{\mathrm{merger}}$, right panel) between matched merger pairs of TNG100-1 and TNG100-1-Dark. In each panel, the grey scale indicates the number of merger events included in each pixel in the figure with timestep of 0.24 Gyr (infall time), 0.16 Gyr (merger time), and 0.18 Gyr (timescale), respectively. The grey dashed line is the diagonal line. In the middle/right panel, the red line shows the best-fit linear relation with the slope fixed to unity, obtained via iteratively reweighted least squares \citep{nelder1972generalized,holland1977robust}, of which the equation is shown in the upper-left corner in the panel. \adr{}The matched mergers have similar infall times, but in general later merger time and therefore longer merger timescales in the 
DMO simulation than in the hydro simulation.}

\label{fig:3 time}
\end{figure}

Fig.~\ref{fig:3 time} shows a one-to-one comparison of $t_\mathrm{infall}$, $t_\mathrm{merger}$, and merger timescale $t_\mathrm{infall}-t_\mathrm{merger}$ for the matched mergers.
%The grey scale in each panel indicates the number of merger events included in each pixel in the figure with timestep of 0.24 Gyr (infall time), 0.16 Gyr (merger time), and 0.18 Gyr (timescale), respectively. %The grey dashed line is the diagonal line.
As shown in the left panel, the matched pairs share similar $t_\mathrm{infall}$, of which 45.7\% have the same infall times, and other 26.3\% have infall times that differ by no more than 0.24 Gyr (by one snapshot). 
The rest of the matched pairs mostly exhibit earlier infall times in the hydrodynamical simulation.
In the middle panel, the differences of $t_\mathrm{merger}$ between the two simulations are much larger compared to that of $t_\mathrm{infall}$. The red line shows the best-fit linear relation with slope fixed to unity. The red line shifts downward by 0.57 Gyr compared to the diagonal gray line, implying on average a delay of mergers in DMO simulation relative to hydrodynamical simulation. The right panel of Fig.~\ref{fig:3 time} presents the comparison of merger timescale $t_\mathrm{infall}-t_\mathrm{merger}$ for the matched merger pairs. The red best-fit line with slope fixed to unity shifts upwards by 0.36 Gyr, indicating a shorter merger timescale in general in the hydrodynamical simulation.

From Fig.~\ref{fig:3 time} we see that in general, the matched mergers in TNG100-1 and TNG100-1-Dark have similar infall times, but differ in much larger ranges for the merger times. Mergers in the DMO simulation typically happen later, and therefore have longer merger timescales, compared with those in the hydro simulation, which is consistent with \citet{xu2025describing} in the halo mass and stellar mass range of interest. We have checked that these results remain similar for mergers in different galaxy mass bins, and for both major and minor mergers individually.

\subsection{Merger orbit comparison for major mergers in massive galaxies\label{subsection orbits}}

In this subsection, for the matched sample of major mergers for massive galaxies in TNG100-1 and TNG100-1-Dark, we first compare in detail their merger orbits, showing two examples of merger pairs, one with distinct merger orbits and one with similar merger orbits. Then we statistically compare the merger orbital type (indicated by collision angle) for the whole sample. 

\begin{figure}[h!]
\centering
\includegraphics[width=0.98\textwidth]{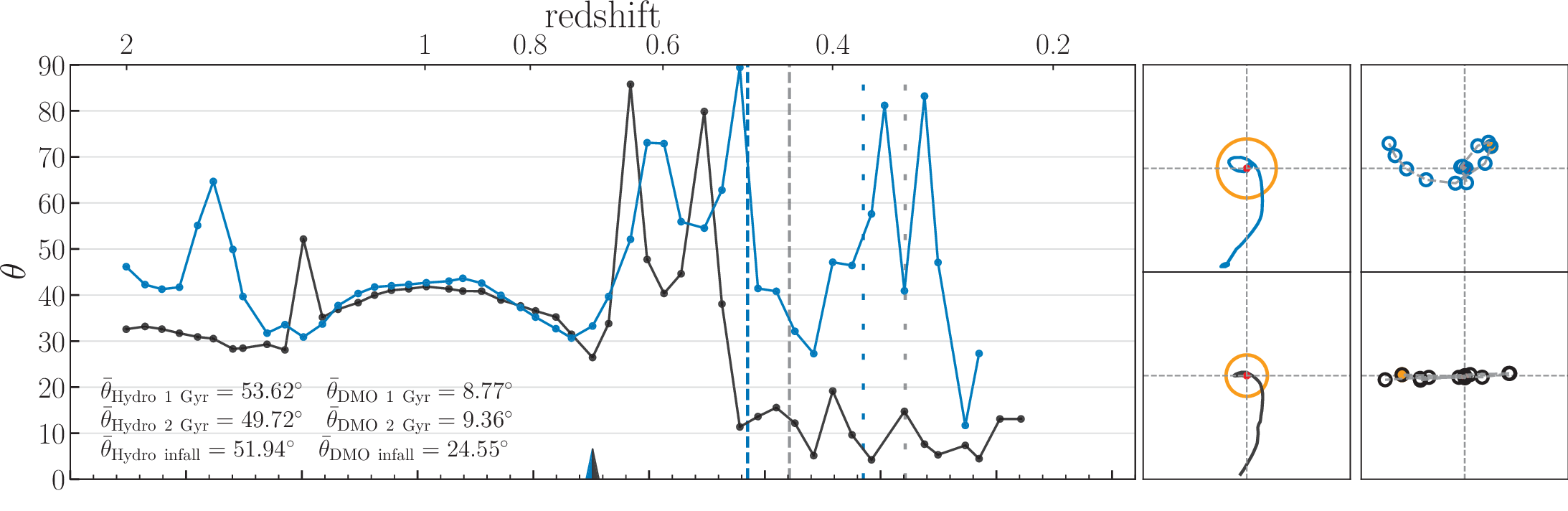}
\includegraphics[width=0.98\textwidth]{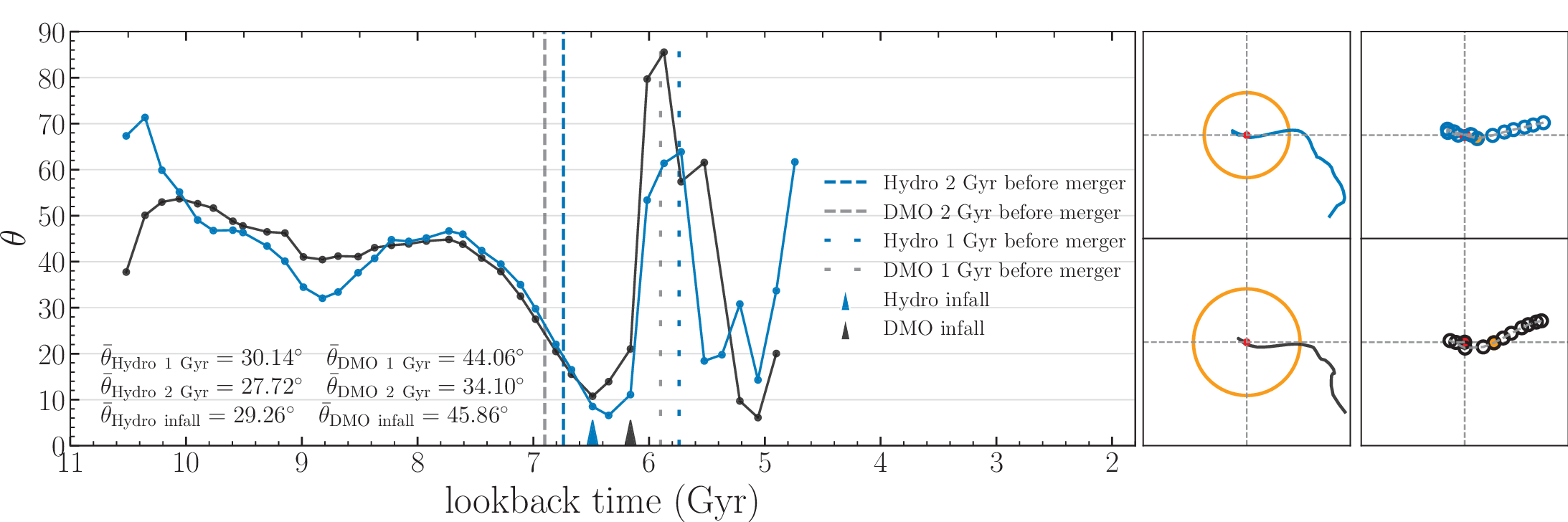}
\caption{Two examples of matched major mergers for massive galaxies in TNG100-1 (blue lines and symbols) and TNG100-1-Dark (black lines and symbols). 
In each row, the left panel shows the evolution of $\theta$ as a function of lookback time, for the merger in the TNG100-1 simulation (blue line) and the one in the TNG100-1-Dark simulation (black line), from $z = 2$ to $t_\mathrm{merger}$. Triangle symbols of corresponding colors along the $x$-axis indicates the infall time $t_\mathrm{infall}$ of satellite galaxy in the merger. 
Two vertical dashed lines of corresponding colors indicate the times 2~Gyr and 1~Gyr before $t_\mathrm{merger}$, respectively. 
The average angles $\bar{\theta}$ from the two simulations are listed at the bottom left corner, when averaging $\theta$ in lookback time intervals of [$t_\mathrm{merger}$+1Gyr, $t_\mathrm{merger}$], [$t_\mathrm{merger}$+2Gyr, $t_\mathrm{merger}$], and [$t_\mathrm{infall}$, $t_\mathrm{merger}$],
%1~Gyr before the merger, 2~Gyr before the merger, and from $t_\mathrm{infall}$ to $t_\mathrm{merger}$, 
respectively. 
In each row, the four small panels on the right show the merger orbits in the TNG100-1 simulation (upper two panels), and in the TNG100-1-Dark simulation (lower two panels), with the central galaxy/subhalo fixed in the center. In each left small panel, the orange circle indicates $R_{200}$ of the central subhalo one snapshot before $t_\mathrm{infall}$, while the solid line shows the trajectory of the satellite galaxy/subhalo relative to the central. The right small panels present the zoomed-in trajectory of the orbits after $t_\mathrm{infall}$, with open circles indicating the satellite positions. The orange filled circle in each right small panel indicates the closest snapshot to the time of 1Gyr before $t_\mathrm{merger}$.}
\label{fig:orbit type}
\end{figure}

In Fig.~\ref{fig:orbit type}, we show the two examples of matched merger pairs. In each row, the left panel shows the evolution of $\theta$ for mergers in the two simulations, where $\theta$ is the acute angle between the relative position vector and the relative velocity vector of the satellite with respect to the central. $\theta$ close to 90$^\circ$ corresponds to spiral-in orbits, while small $\theta$ indicates head-on mergers.
The example shown in the upper row is a typical merger pair with the same $t_\mathrm{infall}$, a bit later $t_\mathrm{merger}$ and therefore a longer merger timescale in DMO simulation than in the hydro. The merger is close to a spiral-in orbit in TNG100-1 (blue lines), but almost a head-on collision in TNG100-1-Dark (black lines). For both mergers of the pair, $\theta$ evolves relatively smoothly at early times, but starts to fluctuate violently after $t_\mathrm{infall}$. After infall, $\theta$ first becomes large, then drops dramatically at the time around 2~Gyr before merger, in both simulations. After that time, $\theta$ remains small in the DMO simulation, but increases again in the hydro simulation. 
The average angle within 1 Gyr before merger is 53.62$^\circ$ in the hydro and 8.77$^\circ$ in the DMO simulation. If averaging the angle in a longer timescale, the differences between the two become smaller.
%gradually stabilizes and begins to reflect the final merger orbits. 

The right small panels in the upper row of Fig.~\ref{fig:orbit type} show the corresponding merger orbits for the example pair, from which we can see the orbital type more directly. The two merger orbits, in general, have similar trajectories before infall. After infall, the influence of the central galaxy becomes evident as the satellite is pulled closer to the center, changing its orbital direction. The orbits of the two mergers start to differ in the last 2 Gyr before merger, and result in a final spiral-in merger orbit in the hydro simulation and a head-on collision in the DMO simulation.

In the bottom row of Fig.~\ref{fig:orbit type}, we give another example of matched merger pairs, in which the two mergers show a relatively similar evolution of $\theta$ and have a similar merger orbital type. Similarly, as seen in the example shown in the upper row, $\theta$ begins to change more dramatically after $t_\mathrm{infall}$, indicating that the influence of the central on the orbit of the satellite galaxy at roughly this time. For this merger pair, the average $\theta$ in the two simulations are similar, for all the three time intervals adopted, with $\theta$ in DMO simulation a bit larger than that in the hydro. %Note that in this example, the merger timescales ($t_{\mathrm{infall}} - t_{\mathrm{merger}}$) in both simulations are shorter than 2~Gyr.
%and the differences between $\theta$ averaged within 1Gyr before the merger is larger than that within 2 Gyr before the merger. 

\begin{figure}[htbp]
\centering
\includegraphics[width=0.9\textwidth]{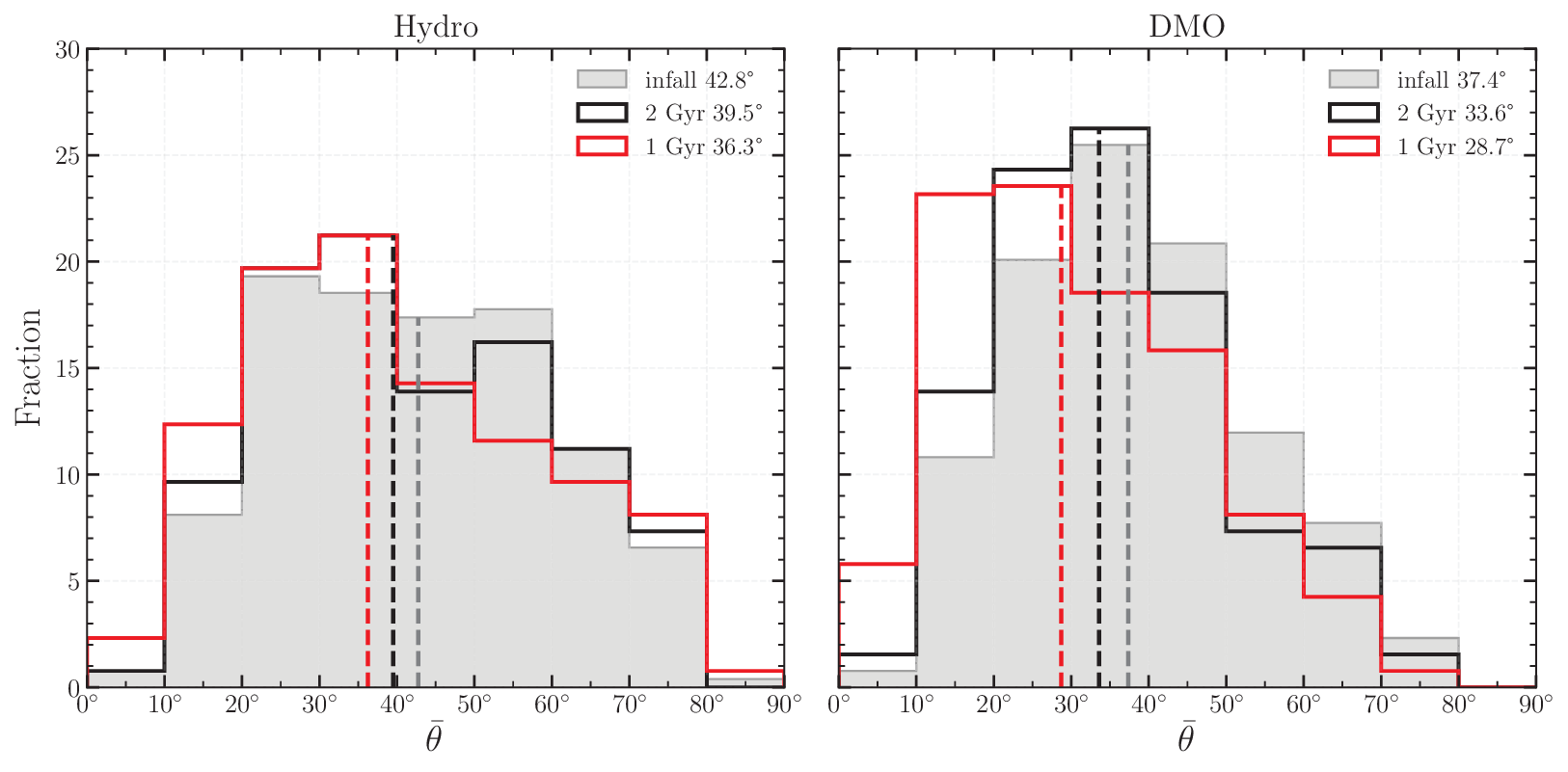}
\caption{Distributions of the collision angle of mergers for the matched major merger pairs in TNG100-1 (left) and in TNG100-1-Dark (right). The red histograms are results of the collision angle measured from 1~Gyr before merger till merger time.
The gray filled and black histograms show respectively the distributions of collision angles averaged starting from 2 Gyr before merger, and starting from infall time. The red, black and grey vertical dashed lines indicate the median values of the corresponding distributions, with the numbers presented in the upper right corner of each panel. In both simulations, with shorter time intervals applied, the distribution of collision angle shifts a bit to the smaller value, while the median value remains larger in the hydro simulation than in the DMO one. }
\label{fig: 3 angle distribution}
\end{figure}

In Fig.~\ref{fig: 3 angle distribution}, we statistically compare, for all the matched major mergers of TNG100-1 (left panel) and TNG100-1-Dark (right panel), the distribution of collision angle of the mergers. The collision angle is defined as the average $\theta$ measured from 1 Gyr before merger till merger time in \citet{zeng2021formation}, and the corresponding results are shown in red histograms. We have also checked the results when calculating the collision angle as the average $\theta$ from 2 Gyr before merger till merger time (black histograms), and from infall time till merger time (gray histograms). While $\theta$ starts to be dramatically affected by the merger after infall time as seen in Fig.~\ref{fig:orbit type}, we aim to check whether different time intervals applied for measuring collision angle would affect the related statistics. 

%\adr{Among all the massive major mergers 13 out of 259 have infall time within 1 Gyr, and 81 out of 259 mergers have infall time within 2 Gyr.} 
Note that for the mergers considered, infall of the satellite can happen within 2 Gyr and even 1 Gyr before the merger time, as can be seen in the example shown in the bottom row of Fig.~\ref{fig:orbit type}, and for the cases with merger timescales less than 2 Gyr, as shown in the right panel of Fig.~\ref{fig:3 time}. For all the 259 matched major mergers in massive galaxies, 81 of them have $t_\mathrm{infall}$ within 2 Gyr before merger, among which 13 have $t_\mathrm{infall}$ within 1 Gyr before merger in at least one simulation. While merger orbits start to change dramatically after $t_\mathrm{infall}$, when considering time intervals from 2 Gyr/1 Gyr before merger to merger time, we only include snapshots after $t_\mathrm{infall}$, which applies to all the following analysis related. Fig.~\ref{fig: 3 angle distribution} shows that with shorter time intervals applied, the distribution of collision angle shifts a bit to the smaller value, in both simulations, with a larger deviation between different histograms in the DMO simulation than in the hydro.

\begin{figure}[htbp]
\centering
\includegraphics[width=0.9\textwidth]{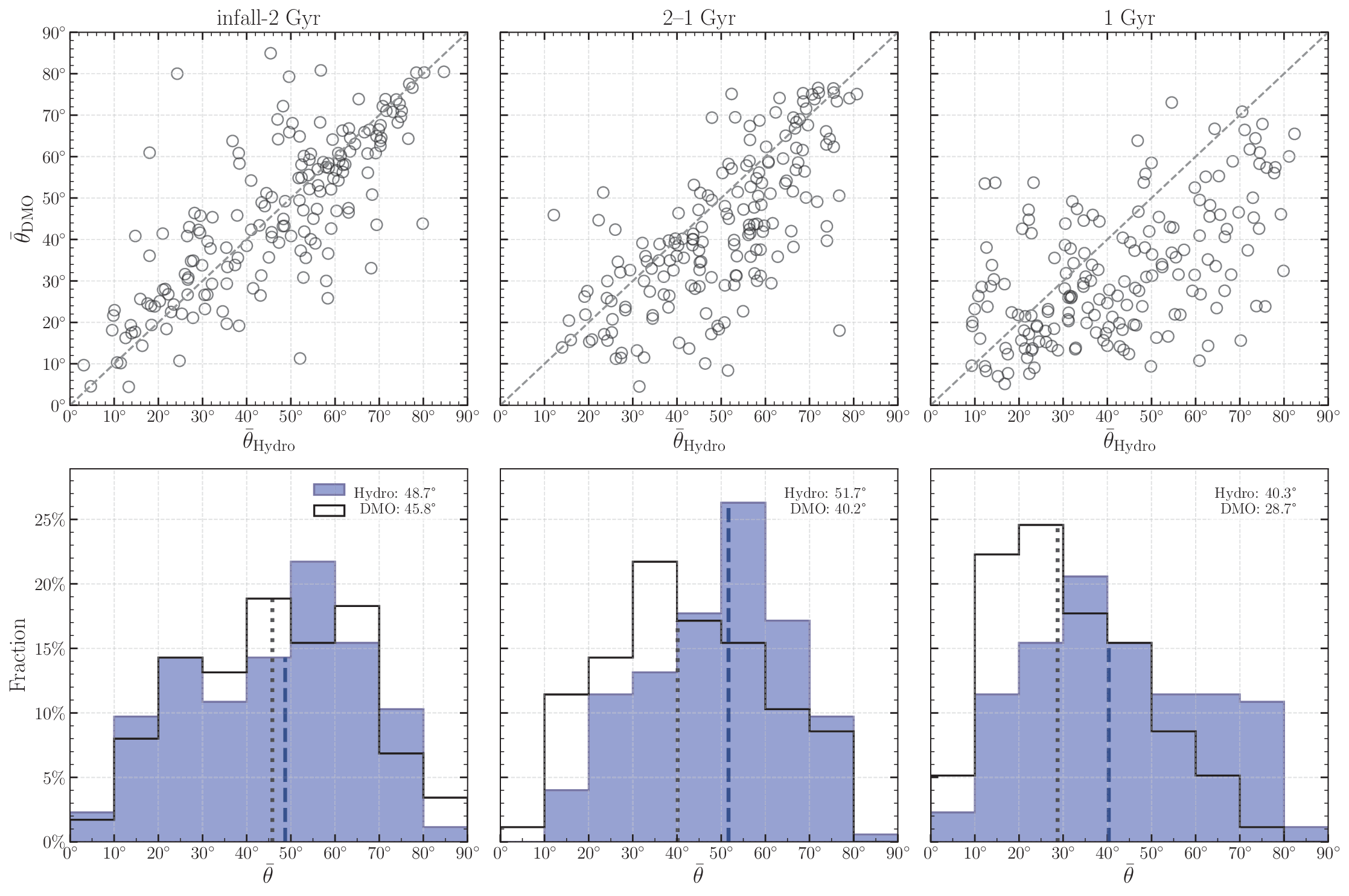}
\caption{For the 175 matched major merger pairs with $t_\mathrm{infall} > 2$ Gyr in TNG100-1 and TNG100-1-Dark, one-to-one comparison of the average orbital angles (upper panels) and the distributions of the average orbital angles (lower panels) in the two simulations. From left to right, results are shown for angles averaged in time intervals of [$t_\mathrm{infall}$, 2 Gyr + $t_\mathrm{merger}$], [2 Gyr + $t_\mathrm{merger}$, 1 Gyr + $t_\mathrm{merger}$], and [1 Gyr + $t_\mathrm{merger}$, $t_\mathrm{merger}$].
%Left column: the average angles between $t_\mathrm{infall}$ and 2 Gyr before $t_\mathrm{merger}$. Middle column: the average angles between 2 and 1 Gyr before merger. Right column: the average angles within 1 Gyr before $t_\mathrm{merger}$. 
The grey dashed lines in the upper panels are the diagonal lines. In the lower panels, the blue filled histograms show results from the hydro simulation while the black histograms show those from the DMO simulation. The blue and black vertical dashed lines in the lower panels indicate the median angles in the hydro and DMO runs, respectively, with the median values presented in the upper right of each panel. In both simulations, after infall, the merger
orbits evolve from more spiral-in to more head-on, with mergers in the DMO simulation evolving much more than those in the hydro simulation.}
\label{fig:3 average angle}
\end{figure}

In Fig.~\ref{fig:3 average angle}, we compare in detail the average orbital angles between the hydro simulation and the DMO simulation, when the angle is averaged in three sequential time intervals, to investigate how the average $\theta$ evolves statistically from $t_\mathrm{infall}$ to $t_\mathrm{merger}$. 175 matched major merger pairs only with $t_\mathrm{infall} > 2$ Gyr are shown in Fig.~\ref{fig:3 average angle}, in order to track the evolution of orbital angle for the same sample in the three time intervals investigated. We have checked that for major merger pairs with $t_\mathrm{infall} < 2$ Gyr, the main results presented below remain similar, with the median value of angle distributions differing a bit. 
%massive major mergers with $t_{timescale} < 2$, shown in Fig.\ref{fig:massive_major_rest}.

In the left column of Fig.~\ref{fig:3 average angle}, after infall till 2 Gyr before merger, the orbits of the matched pairs are statistically similar, with points in the upper left panel scattering randomly around the diagonal line, and distributions as well as median angles shown in the bottom left panel being similar. 
At later stages of the mergers, as shown in the right two columns, the averaged angles are statistically larger in the hydro simulation than in the DMO, with median values well separated. The one-to-one comparison points scatter more towards the lower-right corner in the right upper panel, corresponding to an obvious higher fraction of large angles in the hydro simulation than in the DMO as shown in the right bottom panel. 

%\adb{In Fig.\ref{fig:3 average angle} we only show massive major mergers with $t_{timescale} > 2$, we also check massive major mergers with $t_{timescale} < 2$, shown in Fig.\ref{fig:massive_major_rest}. Between infall and 2 Gyr before merge, the orbits of the matched pairs are in general similar, with points in the upper left panel scattering closely around the diagonal line, and distributions in the bottom left panel being similar. Within 1 to 2 Gyr before merger, the orbits start to become different, with points in the upper middle panel scattering evenly on the both side of the diagonal line and two distinct peeks in the bottom middle panel. After the time of 1 Gyr before merger, the points in the upper right panel asymmetricly scatter around the diagonal line, with more dispersion on the sub-diagonal side, and the peeks in the distribution on the bottom right showing a shift towards larger angle compare hydro to DMO. Among all the bottom panels, the median number of hydro simulation are all lager than DMO case, suggesting orbits in hydro simulation are more spiral-in.}

When looking at distribution of the average orbital angles after infall in the hydro simulation shown in blue filled histograms in the lower panels of Fig.~\ref{fig:3 average angle}, as time evolves, the median value first becomes a bit larger within 2 Gyr and 1 Gyr before merger, then decreases by a large amount within the final 1 Gyr before merger. The distribution scatters the least around the median within 2 Gyr and 1 Gyr before merger, as shown in the middle panels.
As for the average orbital angles in the DMO simulation shown by black histograms, the distribution shifts towards lower angles and the median value decreases as time evolves, with larger changes at later stage than at earlier stage. Within 1 Gyr before merger, the median collision angle is as small as 28.7$^\circ$.

%\adb{For the results from hydro simulation, as time evolves, the average angle becomes smaller, indicating that satellite orbit becomes more head-on as it approaches the central galaxy, as it shown in the bottom panels, the median number within 2 Gyr firstly become smaller then get larger. At earlier times, some head-on orbits have not yet changed, like the one shown in the upper panel of Fig.~\ref{fig:orbit type}, which have a spiral-in orbit when entering the $R_{200}$, but later changed into a head-on orbit. The situation for the DMO simulation is similar. }

The results above show that, in both hydro and DMO simulations, as time evolves, the averaged orbital angles of the major mergers in massive galaxies become statistically smaller, indicating that the merger orbits evolve from more spiral-in to more head-on. For the matched merger pairs, their orbits are similar right after infall in the two simulations, but mergers in the DMO simulation without baryons evolve to much more head-on orbits. Therefore, the presence of baryons in the TNG100-1 simulation results in more spiral-in orbits for major merger in general.

\subsection{Collision angle distribution for less massive galaxies and for minor mergers\label{Collision angle distribution for less massive galaxies and for minor mergers}}

After looking at the merger orbits of matched pairs for major mergers in massive galaxies, we further extend our study to galaxies of lower mass, and also to the matched minor mergers of TNG100-1 and TNG100-1-Dark. In this subsection, the collision angle is calculated within the time interval from 2 Gyr before merger (and after infall) to $t_\mathrm{merger}$. We choose 2 Gyr before merger to retain a larger set of post-infall snapshots, and have checked that the results shown below remain similar if we use 1 Gyr before merger instead.

\begin{figure}[h!]
\centering
\includegraphics[width=1\textwidth]{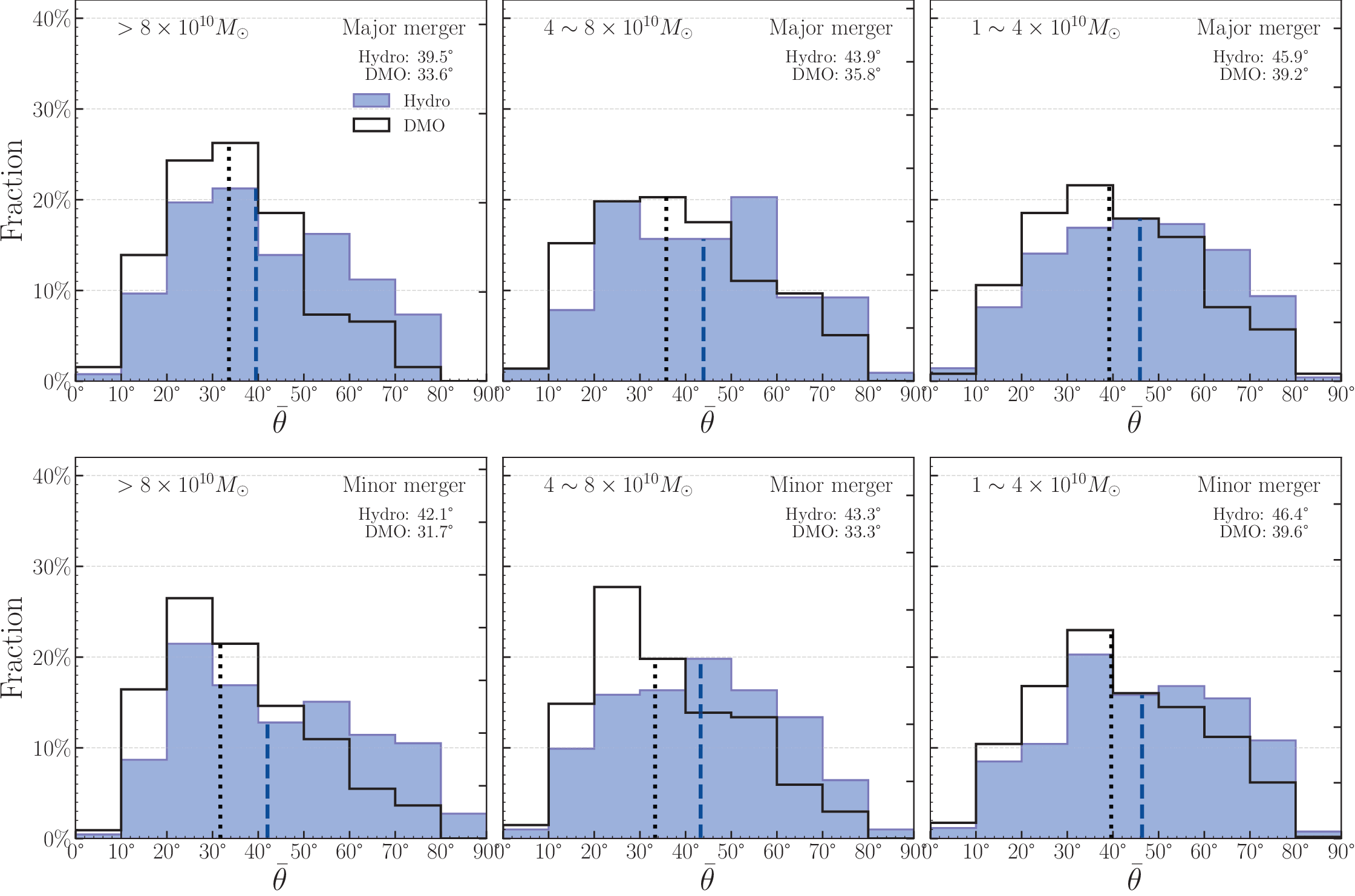}
\caption{The distributions of collision angle for the matched merger samples of TNG100-1 (shaded blue histogram) and TNG100-1-Dark (black histogram) simulations, for major (upper panels) and minor mergers (bottom panels) in galaxies with different stellar mass ranges as indicated in the upper left corner in each panel. The median value of the collision angles in each sample is listed in the right upper corner. For all galaxy masses and merger mass ratios, the mergers in the DMO simulation have smaller median collision angles than in the hydro, indicating that on average, more spiral-in mergers exist when baryons are included in the simulation.
%Panels from left to right correspond to galaxies in the stellar mass ranges of $>8\times10^{10}M_\odot$, $4$–$8\times10^{10}M_\odot$, and $1$–$4\times10^{10}M_\odot$ respectively.
}
\label{fig:distribution 2gyr}
\end{figure}

Fig.~\ref{fig:distribution 2gyr} shows the distributions of collision angles of the matched major mergers (upper panels) and minor mergers (bottom panels), for galaxies in three stellar mass ranges. The upper left panel is the result from the matched major mergers in massive galaxies, which is the sample analyzed in detail in Section~\ref{subsection orbits}. The median collision angles in the two simulations for this sample differ by 5.9$^\circ$. Compared with this sample, for major mergers in less massive galaxies, the distributions concentrate more towards the intermediate angle in the hydro simulation, with increasing median values, by 6.7$^\circ$ in the lowest mass bin studied. In the DMO simulation, the median collision angle also increases with decreasing galaxy mass with a similar amount as in the hydro simulation, with more extended distributions.

%the collision-angle distribution is mass dependent. In the hydro simulation, as the galaxy stellar mass decreases, the fraction of small-angle declines and the overall distribution shifts toward larger angles. The DMO run exhibits the same mass trend, but at fixed stellar mass it systematically retains a higher small-angle fraction than the hydrodynamic run.

For minor merger pairs, as shown in the bottom panels of Fig.~\ref{fig:distribution 2gyr}, compared with the corresponding results of major mergers in each mass range, the general distributions of collision angles are similar in each simulation, with a small deviation of around 0.1$^\circ$ to 4.5$^\circ$ in median values of the angles.
%\zeng{Okay... but why? This result seems very interesting...}
Again, in both simulations, the median collision angle increases with decreasing galaxy mass, by 4.3$^\circ$ in the hydro simulation and 7.9$^\circ$ in the DMO simulation comparing the highest and lowest massive mass bins. In general, the difference in the median angles is larger for various stellar mass ranges, than for different merger mass ratios in a given simulation.

In all panels of Fig.~\ref{fig:distribution 2gyr}, mergers in the DMO simulation have a smaller median collision angle than in the hydro. This shows that on average, more spiral-in mergers exist when baryons are included in the simulation, which applies to all galaxy masses and merger mass ratios. We have also checked that the evolution trend found in Fig.~\ref{fig:3 average angle} for major mergers in massive galaxies still holds for minor mergers and for mergers in galaxies of lower mass, with similar merger orbits after infall in the two simulations evolving both to more head-on orbits and evolving more in the DMO simulations.
%the collision-angle distribution in the hydro simulation becomes progressively shifted toward larger $\theta$ as the stellar mass of the central galaxy decreases, with a corresponding decline in the small-angle fraction. The DMO counterpart exhibits the same mass-dependent trend, but at fixed stellar mass it systematically retains a higher fraction of small-$\theta$ encounters than the hydro run, similar to what we found for major mergers.

%At fixed stellar mass, the hydro run shows that minor mergers shift toward smaller collision angles than major mergers in the two higher mass bins ($>8\times10^{10}M_\odot$ and $4$–$8\times10^{10}M_\odot$), while the lowest mass bin ($1$–$4\times10^{10}M_\odot$) shows little difference between minor and major mergers. The DMO run presents the same tendency in the two higher- mass bins. Minor mergers are slightly more weighted toward small angles than major mergers—again with negligible difference in the lowest mass bin.

%Overall, our results reveal a general trend, the collision angle distribution is primarily governed by the stellar mass of the central galaxy. As stellar mass decreases, both major and minor mergers shift from small to larger with a declining small-angle fraction. In contrast, the variation with merger mass ratio is comparatively modest at fixed mass, the stellar-mass dependence is noticeably stronger than the merger–mass–ratio dependence.

\section{Effect of numerical resolution on orbital type}
\label{section: Comparison in different resolutions}

In both hydro and DMO simulations, identification of galaxies and subhaloes, both central and satellite, depends on the numerical resolution of the simulations, including particle masses and softening lengths \citep{pillepich2018simulating,pillepich2018first}. As a result, the identification of mergers also inevitably depends on the numerical resolution of simulations. In this section, we therefore examine how the results from the previous sections would be affected by resolution, by looking at TNG100-2 and TNG100-3 and the corresponding DMO runs, which have lower resolutions than the TNG100-1 runs analyzed above. At low resolutions, softened potentials and poorly resolved galaxy structure suppress the impact of baryons on halo and galaxy internal structure, normally making the hydro runs more DMO-like \citep{ludlow2020numerical}. However, the detailed resolution effects on convergence are complicated, affected by the choices of the gravitational softening, time-step, force accuracy, initial redshift, particle number \citep{power2003inner,zhang2019optimal}, as well as by the two-body scattering effect \citep[e.g.,][]{ludlow2020numerical,wilkinson2023impact,zeng2024kinematic}.
%the dependence of collision angle on galaxy mass, merger mass ratio as well as the inclusion of baryons as studied in the previous sections remains in lower-resolution simulations. 
%ote that the merger samples analyzed are matched between the corresponding hydro and DMO simulations with the same resolution level. Matching is not applied between different levels, since too few mergers can be found if trying to do so.
We therefore only analyze samples from different resolution levels separately, and compare the overall distribution and median of collision angle between different levels, not aiming to study the detailed resolution effect on specific mergers. 

The same method of selecting matched merger pairs as used for TNG100-1 runs is applied to both TNG100-2 and -3 runs.
%Similar as for mergers in TNG100-1, we apply the selection criterion and matching method described in section~\ref{Matching} for galaxies and mergers in TNG100-2 and -3 and the corresponding DMO simulations. 
As described in Section \ref{merger samples}, mergers of galaxies are classified into different mass bins according to their stellar masses in TNG100-1, and of subhaloes more massive than \(8.9 \times 10^{9}\, h^{-1} M_{\odot} \) in TNG100-1-Dark are selected in the analysis above. To be consistent with the mass limit, we classify mergers of galaxies in the same way as in TNG100-1, and select mergers of galaxies containing more than 125 particles in TNG100-2-Dark, and more than 16 particles in TNG100-3-Dark. The total numbers of mergers selected for different galaxy mass bins in the hydro simulations are listed in Table~\ref{tab:examples 2 and 3}, and the numbers in brackets are the number of mergers after matching with the corresponding DMO simulations. Compared with the numbers listed in Table~\ref{tab:examples 2 and 3}, in all mass and mass-ratio bins, merger numbers and matched merger numbers decrease with decreasing resolutions, due to fewer galaxies/subhaloes identified.

\begin{figure}[h!]
\centering
\includegraphics[width=0.99\textwidth]{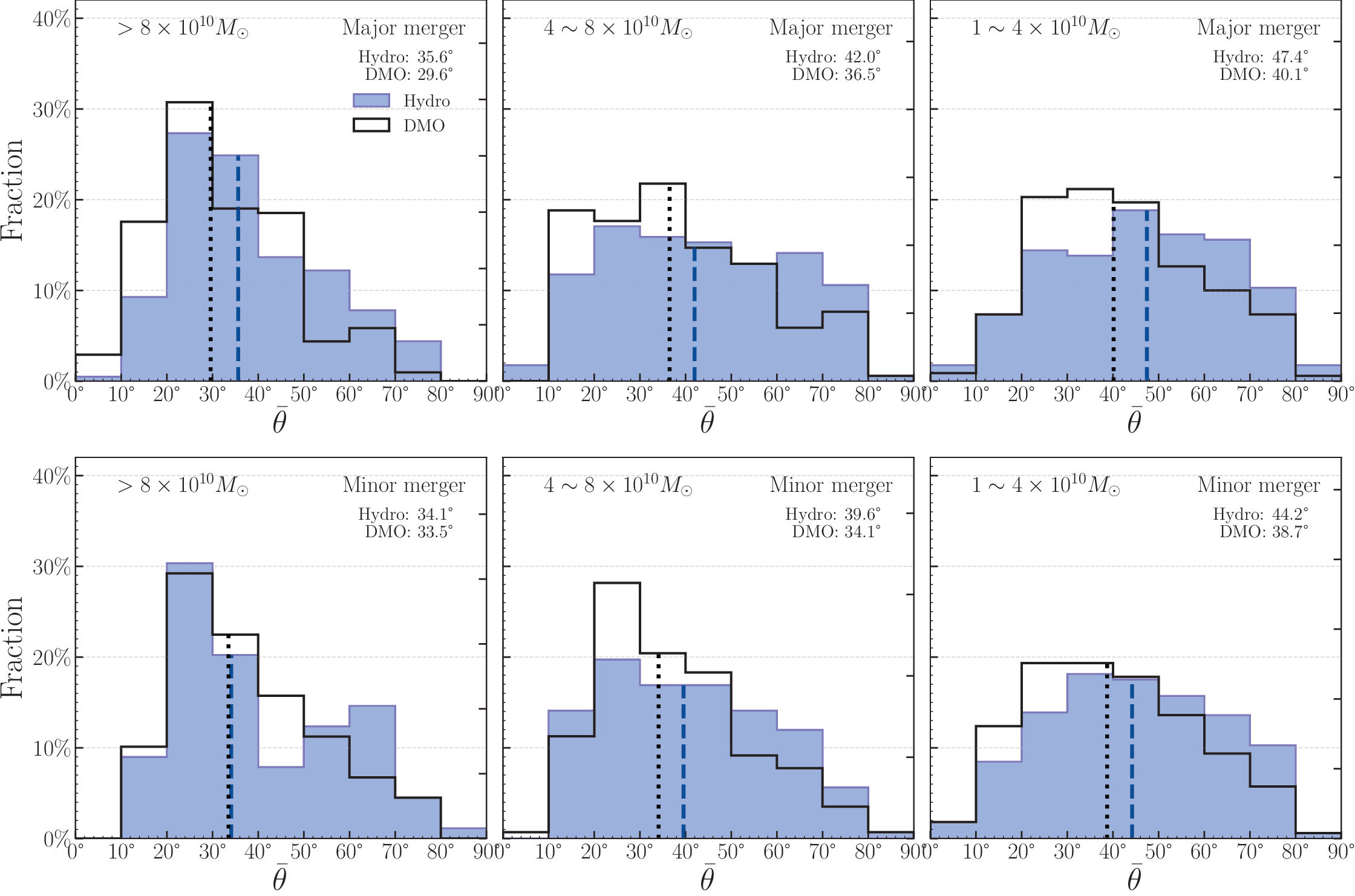}
\caption{The same as Fig.~\ref{fig:distribution 2gyr}, but for matched merger samples selected from TNG100-2 and TNG100-2-Dark. With lower resolutions, merger orbits in the hydro simulations are less
spiral-in, while merger orbits in the DMO simulations are roughly not affected.}
\label{fig:distribution_2gyr-2}
\end{figure}

\begin{figure}[h!]
\centering
\includegraphics[width=0.99\textwidth]{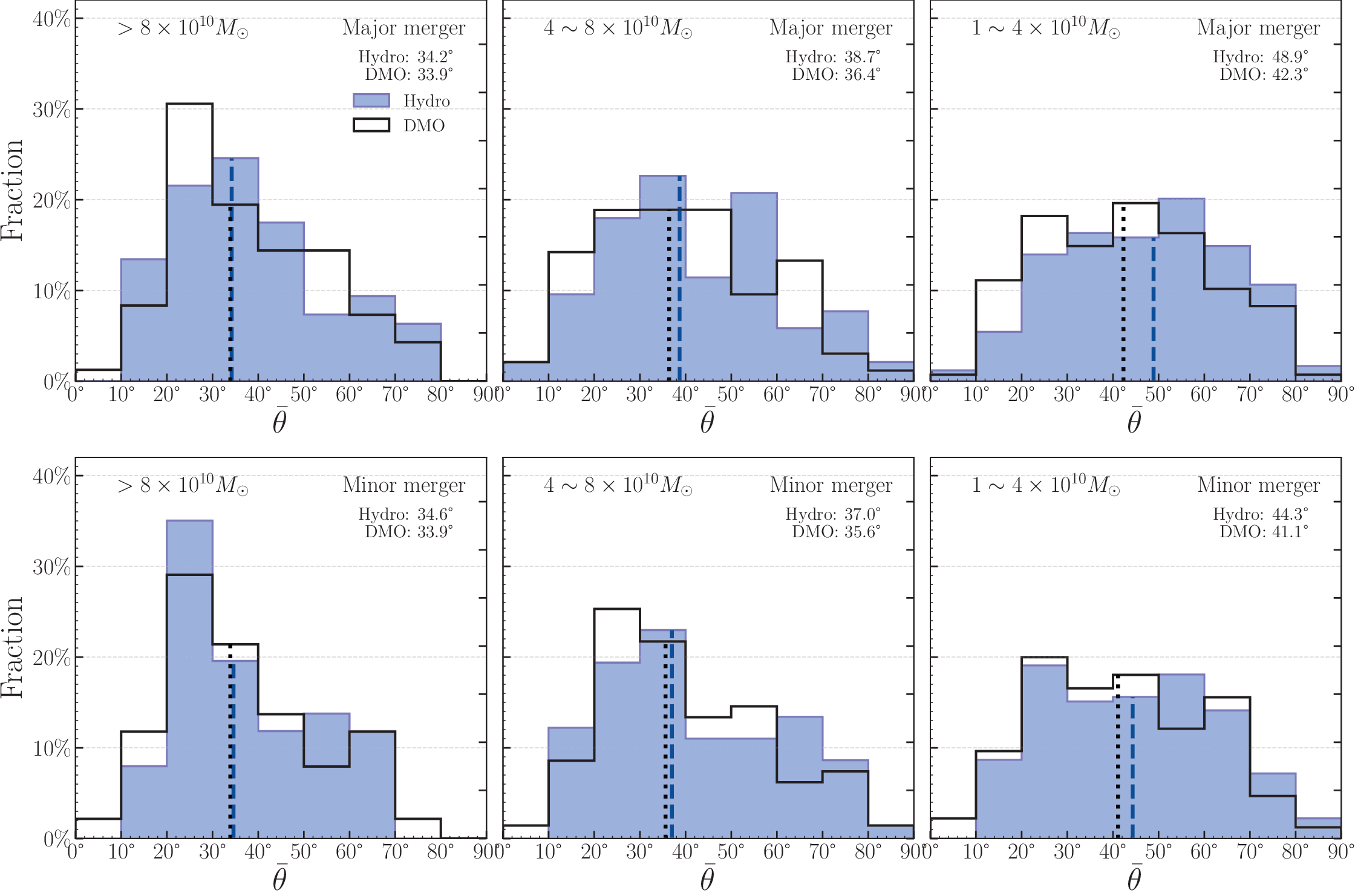}
\caption{The same as Fig.~\ref{fig:distribution 2gyr}, but for matched merger samples selected from TNG100-3 and TNG100-3-Dark. %Similar as Fig.\ref{fig:distribution 2gyr}, the distributions of the collision angle, but for matched merger samples selected from -3. The shaded blue histograms show results from -3 hydro simulations, while black histograms show results from -3 DMO simulations.
} 
\label{fig:distribution_2gyr-3}
\end{figure}

\begin{table}[h]
\centering
\caption{
The total numbers of major and minor merger events selected in TNG100-2 and -3 simulations, for galaxies of different stellar masses at present day. The numbers in brackets are the numbers of mergers after one-to-one matching with the TNG100-2 and -3 DMO simulations respectively.}
\label{tab:examples 2 and 3}
\begin{tabular}{lcccc}
\toprule
 &  & Massive  & MW & Less\\
\midrule
TNG100-2 &Major & 351 (205) & 295 (170) & 654 (340) \\
 & Minor & 229 (89) & 239 (142) & 600 (331) \\
\midrule

TNG100-3 & Major & 180 (97) & 184 (105) & 412 (207) \\
 & Minor & 117 (52) & 127 (82) & 337 (199) \\
\bottomrule
\end{tabular}
\end{table}

In Fig.~\ref{fig:distribution_2gyr-2} and Fig.~\ref{fig:distribution_2gyr-3}, we present the distributions of collision angles for mergers matched between the hydro and DMO simulations of TNG100-2, and TNG100-3, respectively. Compared with the results of TNG100-1 shown in Fig.~\ref{fig:distribution 2gyr}, for the samples in the same range of galaxy mass and merger mass ratio, the distributions are in general similar between different levels, more for low mass galaxies than for massive galaxies. When looking at the median values of the distributions of different levels, in the hydro simulations, median angles are smaller with lower resolutions in the two massive mass bins for both minor and major mergers. In the DMO simulations, the median angles in general change much less and roughly remain similar with different resolutions. As a result, the differences in median angles between hydro and DMO simulations become smaller for lower resolutions, with the fact still holds that mergers in the hydro simulations have a larger median collision angle than in the DMO, as has been seen in Fig.~\ref{fig:distribution 2gyr}.

Comparing Fig.~\ref{fig:distribution 2gyr} with Fig.~\ref{fig:distribution_2gyr-2} and Fig.~\ref{fig:distribution_2gyr-3}, we find that merger orbits in the hydro simulations are less spiral-in with lower resolutions, and merger orbits in the DMO simulations are roughly not affected by resolution.
Therefore, while including baryons in the simulation always statistically makes the merger orbit a bit more spiral-in, the effect is less obvious with lower resolutions.

% , and comparing the corresponding median values, showing a a systematic shift in the hydro simulations toward larger values. The median values for -1, -2, and -3 are broadly similar, and the Hydro medians are consistently higher than the DMO medians. In -2, the Hydro–DMO medians are generally closer than in -1; the only exceptions are the massive-major and less-massive-major bins, where the Hydro–DMO offsets are larger by 0.1° and 0.7°, respectively. In -3, all Hydro–DMO medians are closer than in -1; relative to -2, the major–minor category shows a 0.1° larger offset, indicating that the lower resolution simulations show smaller differences.

%Fig.5 and Fig.6 indicate that even at lower resolution the baryonic effect remains the same, leading to a systematic shift of the collision angles in the hydro simulations toward larger values compared to the DMO counterparts. At the same time, the distributions of collision angles across different mass ranges and merger categories are also highly consistent with those in TNG100-1. However, as the resolution decreases, the differences between hydro and DMO runs become smaller, indicating that the baryonic effects are more pronounced in higher-resolution simulations.

\section{Discussion and conclusions}
\label{section: Discussion and conclusions}

Using the TNG100 hydrodynamical simulations and their dark-matter-only counterparts, in this work we perform a detailed investigation and comparison of the merger orbits for matched merger pairs across different resolution levels. Our analysis includes both major and minor mergers in galaxies with a range of present-day stellar masses in the hydro simulation: the Milky Way-mass galaxies with $M_{\star} = 4-8 \times 10^{10} {M}_{\odot}$; less massive galaxies with $M_{\star} = 1-4 \times 10^{10} {M}_{\odot}$; and more massive galaxies with $M_{\star} > 8 \times 10^{10} {M}_{\odot}$, together with their dark-matter-only counterparts.

% Using TNG100 hydrodynamical simulations and their corresponding dark-matter-only simulations, we investigate and compare in detail the merger orbits of the matched merger pairs in different simulations of different resolution levels. Matched merger pairs are analyzed for both major and minor mergers, and in galaxies of various present-day stellar masses. 

%we compare in detail the merger orbits in matched merger samples of hydrodynamical TNG simulations and the corresponding pure dark matter TNG-Dark simulations, to investigate how the existence of baryons affects the orbital types of galaxy mergers. 
%In addition, while in simulations numerical resolution is a critical factor influencing the identification of structures and merger dynamics \citep{knebe2011haloes, behroozi2012rockstar, onions2012subhaloes}, we also compare mergers in TNG simulations of different resolutions to study the effects of resolution on merger orbital type.

By comparing all the matched merger samples for massive galaxies in TNG100-1 and TNG100-1-Dark, we find that the matched mergers have similar infall times in the two simulations, but have statistically later merger times in the DMO one. As a result, mergers in the hydro simulation exhibit relatively shorter merger timescales on average, indicating that the presence of baryons accelerates a bit the merger processes statistically.
%statistically, the infall times are identical in most cases between the two simulations, whereas the merger times are systematically later in the hydro runs. As a result, mergers in the DMO simulations generally exhibit relatively shorter merger timescales. 

For the matched major merger samples of massive galaxies in TNG100-1 and TNG100-1-Dark, we analyze in detail their merger orbits and compare the evolution of the collision angle (the averaged orbital angle) which quantitatively represents the orbital type. Right after infall, the merger orbits are statistically similar in the two simulations. As time evolves, the averaged orbital angles in both simulations become statistically smaller, reflecting more head-on orbits at later stages. The decreasing of the angle with time is much larger in the DMO simulation than in the hydro. After the time of 2 Gyr before merger, the collision angle remains in general smaller in the DMO simulation than in the hydro, indicating that the presence of baryons results in more spiral-in orbits for major mergers. This result also applies to all galaxy masses and merger mass ratios investigated, with mergers in lower-mass galaxies on average having more spiral-in orbits.

Analyzing the matched mergers in the two lower-resolution simulations TNG100-2/TNG100-3 and their corresponding DMO simulations, together with the results from TNG100-1, we find that with lower resolutions, merger orbits in the hydro simulations are less spiral-in, while those in the DMO simulations roughly remain unaffected. While including baryons in the hydro simulation always statistically makes the merger orbit a bit more spiral-in than in the DMO, the difference between the two types of simulations is in general smaller with lower resolution levels, indicating that the effect is less obvious with lower resolutions.
%with the distributions systematically shifted toward larger angles, by [5.9°,10.4°] for the median value. The baryonic effects on collision angle are more obvious in higher-resolution simulations, by [0.6°,7.4°] in -2 and [0.3°,6.6°] in -3 for the median value.

%Our results demonstrate that with baryons included in the TNG simulations, merger timescales are statistically larger, and merger orbits are more spiral-in, compared with the corresponding DMO simulations. In all stellar mass ranges and mass ratio ranges investigated, the differences of collision angles between two set of simulations various by 14.6\%-24.7\% percent. The difference is in general not large, and the general distribution of collision angle in hydro simulation remains unchanged in DMO simulation. Nevertheless, with deviations existing, this difference should be considered when trying to link galaxy morphology to mergers and merger orbit for halo-based models or semi-analytic models of galaxy formation which are built on DMO simulations.

Our results demonstrate that with baryons included in the TNG simulations, merger timescales are statistically a bit shorter, and merger orbits are on average more spiral-in, compared with the DMO simulations. Nevertheless, the differences in collision angles in the matched pairs are not large in general. Therefore, 
the strong dependence of galaxy post-merger morphology on collision angle found based on TNG100-1 simulation in \citet{zeng2021formation} can shed light and be applied to halo-based models and semi-analytic models of galaxy formation, which is based on DMO simulations. The latter attempt will be studied in a following work (Xie in preparation). Also, as discussed in Section \ref{sect:intro}, in other hydro simulations with different descriptions of baryon processes applied, the dependence of galaxy morphology on merger properties could be quite different. Therefore, it is necessary to check whether the results of \citet{zeng2021formation} still hold in simulations other than the TNG, which will be studied in the following work (Gan et al. in preparation).

\begin{acknowledgements}
We thank Shihong Liao and Qianyu Gan for helpful discussions. We acknowledge the supports from the National Natural Science Foundation of China (grants No.12588202, No.12473015), 
the National SKA Program of China (No.2022SKA0110201), the National Key Research and Development Program of China (No.2023YFB3002501), and the Strategic Priority Research Program of Chinese Academy of Sciences, (grant No. XDB0500203).
%and K.C. Wong Education Foundation.
\end{acknowledgements}

\appendix
\section{Collision angle distribution in major merger samples}                  %%appendicial material is supported

As described in Section \ref{merger samples}, the criteria for selecting mergers used in this work have some slight differences compared to the ones applied in \citet{zeng2021formation}. Specifically, the definition of galaxy stellar mass, merger mass ratios are not exactly the same. In addition, major or minor mergers that experience another major merger within 1~Gyr before or after the event are excluded. We have checked that the main results of \citet{zeng2021formation} still hold when applying these updates to the major merger sample of TNG100-1 massive galaxies therein. 

%Throughout the paper, we focus on the collision angle as defined in \citet{zeng2021formation} to represent the orbit type of galaxy mergers. The collision angle is defined as the average acute angle, measured within 1~Gyr of each snapshot, between the relative position vector and the relative velocity vector of the satellite galaxy with respect to its central galaxy.

\begin{figure}[htbp]
\centering
\includegraphics[width=0.45\textwidth]{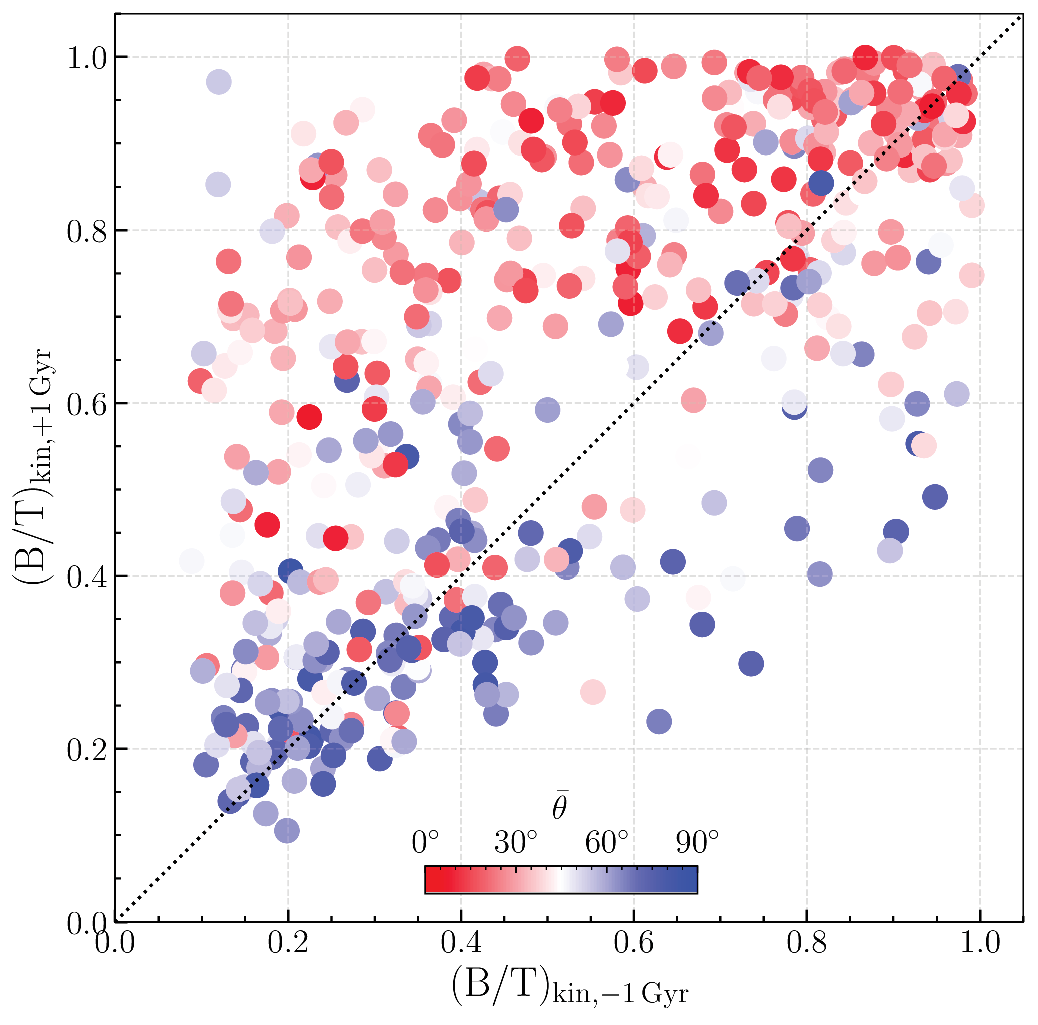}
\includegraphics[width=0.45\textwidth]{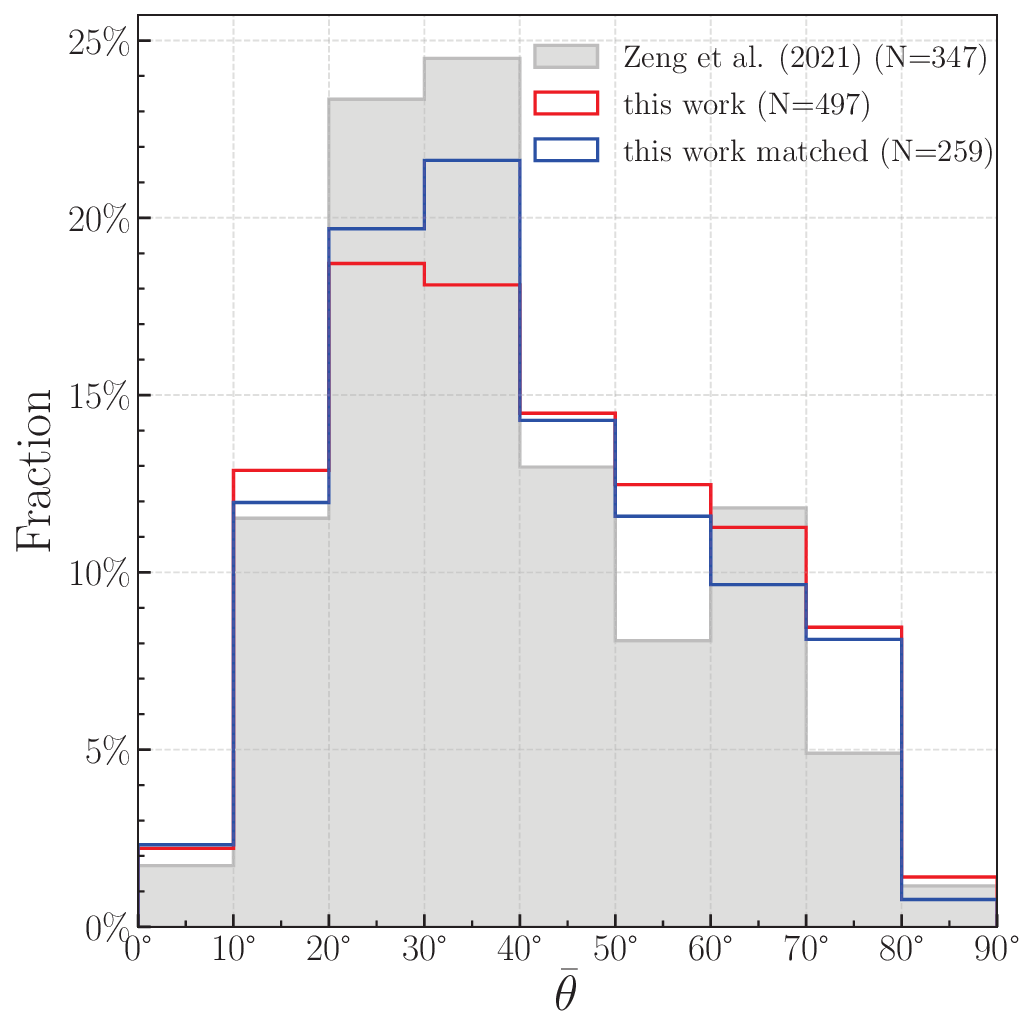}
\caption{Left panel: the distributions of collision angle in different samples. The gray histogram is the result of the major merger sample for massive galaxies in TNG100-1 as used in \citet{zeng2021formation}. The red line indicates the result of the major merger sample with updated criteria as used in this work, and the blue line is for the merger sample after matching with the TNG100-1-Dark simulation. The number of mergers in each sample is shown in bracket in the upper right corner. Right panel: same as in Fig.7 of \citet{zeng2021formation}, but for the major merger sample for massive galaxies with updated criteria as used in this work, as described in detail in Section \ref{merger samples}.}
\label{fig:compare with previous work}
\end{figure}

In Fig. \ref{fig:compare with previous work}, the left panel is the same as Fig.7 in \citet{zeng2021formation}, but for the sample with the updated criteria. The main result still holds, that there exists a strong dependence of remnant morphology on the
orbit type (collision angle) of major mergers.

The right panel of Fig. \ref{fig:compare with previous work} shows the distributions of collision angles, in the original sample selected in TNG100-1 by \citet{zeng2021formation}, the TNG100-1 sample with updated criteria as used in this work, and the sample after additionally matching with the TNG100-1-Dark simulation.
The distributions of the three samples all peak around 35$^\circ$. The result of the original sample of  \citet{zeng2021formation} has another small peak around 65$^\circ$, while the distributions are smoother around this value for the other updated samples. The difference in the distributions of the original and updated samples is mainly due to a different stellar mass definition. Nevertheless, the general distribution of collision angle remains similar in the updated samples.

%\begin{figure}[htbp]
%    \centering

%    % 左图
%    \begin{subfigure}[t]{0.65\textwidth}
%        \centering
%        \includegraphics[width=\textwidth]{massive_major_1-2gyr.eps}
%        \caption{Same as Fig.\ref{fig:3 average angle}, but for massive major merger with $2 > t_\mathrm{merger} > 1$. Left column: the average angles between $t_\mathrm{infall}$ and 1 Gyr before $t_\mathrm{merger}$. Right column: the average angles 1 Gyr before $t_\mathrm{merger}$.}
%        \label{fig:massive_major_1-2gyr}
%    \end{subfigure}
%    \hfill
%    % 右图
%    \begin{subfigure}[t]{0.33\textwidth}
%        \centering
%        \includegraphics[width=\textwidth]{massive_major_t_less_than_1.eps}
%        \caption{Same as Fig.\ref{fig:3 average angle}, but for massive major merger with $1 > t_\mathrm{merger}$.}
%        \label{fig:massive_major_t_less_than_1}
%    \end{subfigure}
%
%    \caption{Distribution of massive major merger with $t_\mathrm{merger} < 2$.}
%    \label{fig:massive_major_rest}
%\end{figure}

%\section{This shows the use of appendix}

\bibliographystyle{raa} 
\bibliography{bibtex}
%\begin{thebibliography}{99}
%% you can type \apj for ApJ, \aap for A&A, \apss for Ap&SS, etc. Please consult
%% the macro chjaa.cls. You can also find them in aasguide.tex (AASTeX for ApJ, AJ, PASP)
%% Please follow the format of ChJAA's reference list

%\end{thebibliography}

\label{lastpage}

\end{document}